\documentclass[aoas,MSNbibl,nameyear,dvips]{arximspdf}
\usepackage{dcolumn}
\usepackage{graphicx}

\doi{10.1214/13-AOAS614F} 
\referstodoi{10.1214/12-AOAS614}
\volume{7}
\issue{4}
\pubyear{2013}
\firstpage{1891}
\lastpage{1894}

\makeatletter
\newcolumntype{d}[1]{D{.}{.}{#1}}

\makeatother

\begin{document}
\begin{frontmatter}

\title{Discussion of ``Estimating the historical and future
probabilities of large terrorist events'' by Aaron Clauset and Ryan Woodard}
\runtitle{Discussion}\vspace*{-3pt}

\begin{aug}
\author[a]{\fnms{Qiurong} \snm{Cui}\ead[label=e1]{cui@stat.wisc.edu}},
\author[a]{\fnms{Karl} \snm{Rohe}\corref{}\ead[label=e2]{karlrohe@stat.wisc.edu}}
\and
\author[a]{\fnms{Zhengjun} \snm{Zhang}\ead[label=e3]{zjz@stat.wisc.edu}}
\runauthor{Q. Cui, K. Rohe and Z. Zhang}
\affiliation{University of Wisconsin-Madison}
\address[a]{Department of Statistics\\
University of Wisconsin\\
Madison, Wisconsin 53706\\
USA\\
\printead{e1}\\
\phantom{E-mail:\ }\printead*{e2}\\
\phantom{E-mail:\ }\printead*{e3}}
\pdftitle{Discussion of ``Estimating the historical and future
probabilities of large terrorist events'' by Aaron Clauset and Ryan Woodard}
\end{aug}

\received{\smonth{8} \syear{2013}}
\revised{\smonth{8} \syear{2013}}


\vspace*{-6pt}
\end{frontmatter}

\textit{First and foremost, we commend the authors for their creative
and original investigation. Although this comment will focus on
methodological concerns, we note that the conclusions in Clauset and Woodard
\textup{(\citeyear{clauset})} rely on several strong modeling assumptions. For example, the
number of deaths in a terrorist incident is an independent draw from
some unknown probability distribution that is fixed in space and time.
Readers who are more interested in how Clauset and Woodard \textup{(\citeyear{clauset})} contribute to
the discussion on foreign policy and/or national security should note
that, in general, the advantage of statistical modeling is not
necessarily that the solutions are precise, but rather that all
assumptions are made explicit. Given the politically charged nature of
this problem, we are wary of the assumptions (and thus conclusions) in
the paper.}\vspace*{4pt}

Many of the inferences in \citet{clauset} (hereafter referred to as CW)
rely on the bootstrap to measure the uncertainty in their statistical
estimators. Similarly, other applied papers in the extreme value
literature have relied on the bootstrap [e.g., \citet{zboot}]. However,
there are many scenarios in which the bootstrap can fail. Both \citet
{resnick} and \citet{hall97} discuss some of these problems in the
context of heavy-tailed distributions. In this discussion we provide a
brief simulation that illustrates when the bootstrap succeeds and when
it fails in the settings of CW.


This comment investigates the following question under three relevant models:

\begin{quote}If the bootstrap is used to create a (nominally) 90\%
confidence interval, will this interval actually cover the true
parameter in 90\% of experiments?
\end{quote}

\begin{figure}[b]

\includegraphics{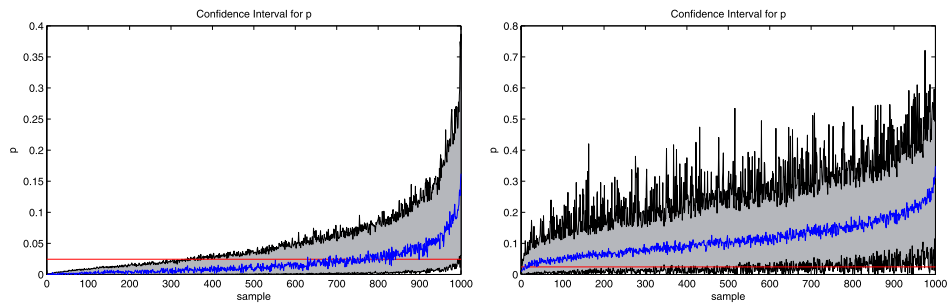}

\caption{One thousand confidence intervals for $p$ under the mixed
power-law distribution with $x_{\min} = 10$ (on left) and GPD with
$x_{\min}$ estimated (on right). These confidence intervals have poor
coverage properties for the true value of $p$ (represented as a red
line). Each confidence interval is computed from a sample of $n=1000$
data points and 1000 bootstrap samples. The order of the intervals is
shuffled so that $\hat{p}$ (blue line) is increasing.}
\label{biasedcase}
\end{figure}

The first simulation model is the power law distribution with $\alpha=
2.4$ supported on $[10,\infty)$. The second simulation model comes from
\citet{clausetPower}, a\vadjust{\goodbreak} paper that CW cite to justify their method of
estimating $x_{\min}$. To sample a point $X_i$ from this model (which
we will refer to as the mixed power-law model), sample an observation
$Y_i$ uniformly at random from the RAND-MIPT data [\cite{rand}] and
sample an observation $Z_i$ from power law with $\alpha= 2.4$
(corresponding to the estimate in CW) and supported on $[10, \infty)$. Then,

\begingroup
\abovedisplayskip=6.7pt
\belowdisplayskip=6.7pt
\[
X_i = \cases{
Y_i, & \quad$\mbox{if $Y_i < x_{\mathrm{min}} = 10$}$,
\vspace*{1pt}\cr
Z_i, & \quad$\mbox{o.w.}$}
\]

To investigate whether the bootstrap techniques in CW are sensitive
to model misspecification error, the final simulation model is the
Generalized Pareto Distribution (GPD) \citet{coles01}.
The GPD distribution is specified by three parameters: location $u$,
scale $\sigma$ and shape $\xi$. The cumulative distribution function of
the GDP distribution is
%
%
\begin{equation}
\label{GPD_CDF} F_{u,\sigma,\xi}(x)=\cases{ %
\displaystyle 1- \biggl(1+\frac{\xi(x-u)}{\sigma} \biggr)^{-1/\xi}, & \quad$\xi\neq0$,
\vspace*{1pt}\cr
\displaystyle 1-\exp \biggl(-\frac{x-u}{\sigma} \biggr), &\quad
$\xi=0.$}
\end{equation}
When $\xi>0$, the GPD distribution is regularly varying at $\infty$
with index $-1/\xi$ and thus tail equivalent to the power-law
distribution with $\alpha=\xi^{-1}+1$.
Therefore, the GPD distribution can be considered as a perturbed
power-law distribution. For simplicity, we set $u=0, \sigma=1$ in the
following simulation study.
For comparison purposes, $\xi$ is set to be $1/(\alpha-1)=1/1.4$ for
$\alpha= 2.4$.

In each of 1000 runs of the experiment, we sample $n=1000$ data points
from each of the simulation models, and we use the code from CW to (1)
fit the power-law distribution and (2) compute the (nominally) 90\%
bootstrap confidence intervals for $\alpha$ and $p$, the probability of
observing at least one catastrophic event. For each of the three
models, we run the estimation two ways: First, with $x_{\min} = 10$
given to the algorithm and, second, where the algorithm estimates
$x_{\min}$. Finally, each bootstrap confidence interval is inspected to
see if it contains the true values of $\alpha$ and $p$ (see next
paragraph for a discussion on computing the true value of~$p$).
In total, this creates six different simulation setups.
The simulation results are given in Table~\ref{default} and Figure~\ref{biasedcase}.

To compute the true value of $p$, we follow the definition in CW:
%
%
\begin{equation}
\label{eq:p} p=1 - E \bigl[P(X < \mathrm{cat}| X > x_{\min})^{n_{\mathrm{tail}}} \bigr],
\end{equation}\endgroup
where $n_{\mathrm{tail}}\sim\mathrm{Binomial}(n=1000,p_{\mathrm{tail}})$,
and $\mathrm{cat}$ is the size of 9/11 (2749). In all six simulation setups,
this value of $p$ is approximately the probability that the maximum of
$n=1000$ draws is greater than $\mathrm{cat}$.\footnote{In fact, for $n=1000$,
they are equal in the first four decimal places.}\vadjust{\goodbreak}

\begin{table}
\caption{For each of the three simulation models, $x_{\min}$ is either
given as 10 or estimated (est.) by the algorithm. The table presents
\textup{(a)} the estimated bias in coverage probabilities for the (nominally) 90$\%$ bootstrap confidence intervals and \textup{(b)} the median width of the
confidence intervals}\label{default}
\begin{tabular*}{\textwidth}{@{\extracolsep{4in minus 4in}}lcd{3.1}d{3.1}cc@{}}
\hline
&&\multicolumn{2}{c}{\mbox{\textbf{Coverage bias (\%)}}} & \multicolumn
{2}{c@{}}{\mbox{\textbf{Width of CI}}} \\[-6pt]
&&\multicolumn{2}{c}{\hrulefill} & \multicolumn
{2}{c@{}}{\hrulefill} \\
\multicolumn{1}{@{}l}{\textbf{Model}} & \multicolumn{1}{c}{$\bolds{x_{\min}}$} &
\multicolumn{1}{c}{$\bolds{\alpha}$}&\multicolumn{1}{c}{$\bolds{p}$}  &\multicolumn{1}{c}{$\bolds{\alpha}$} &\multicolumn{1}{c@{}}{$\bolds{p}$}\\
\hline
{PL}, $\alpha=2.4$ & 10 & -1.0 & -0.9 &0.1 & 0.2\phantom{0}\\
{PL-Mix}, $\alpha=2.4$ & 10 & -28.9 & -25.3  &0.7 &0.03
\\
{GPD}, $\xi=1/1.4$ & 10 & -0.6 & 2.4 & 0.6 & 0.1\phantom{0} \\[3pt]
{Power Law}, $\alpha=2.4$ & \mbox{est.} & 1.5 & 0.9
& 0.2 & 0.3\phantom{0}\\
{PL-Mix}, $\alpha=2.4$ & \mbox{est.} & 0.5 & 0.4  &
0.8 & 0.1\phantom{0} \\
{GPD}, $\xi=1/1.4$ & \mbox{est.} & -11.6 & -18.4  &
0.5 & 0.2\phantom{0}\\
\hline
\end{tabular*}
\end{table}

In the power-law and mixed power-law distributions $x_{\mathrm{min}}=10$
is the lower end of the power-law distribution for the tail model.
Using equation (\ref{eq:p}), $p= 0.3194$ under the power law and $p=0.0244$
under the mixed power law.
Under the GPD distribution an exact $x_{\mathrm{min}}$ is not available
for computing $p$. We therefore applied an analogy of
Kolmogorov--Smirnov criteria to minimize the maximum distance of the GPD
distribution above $x_{\mathrm{min}}$ and power-law distribution, that
is, we find $(x_{\mathrm{min}},\alpha)$ that minimize
\[
f(x_{\mathrm{min}},\alpha) = \max_{x: x>x_{\mathrm{min}}} \biggl\vert \biggl(
\frac{\sigma+\xi(x-u)}{\sigma+\xi(x_{\mathrm{min}}-u)} \biggr)^{-1/\xi
}- \biggl(\frac{x}{x_{\mathrm{min}}}
\biggr)^{1-\alpha} \biggr\vert.
\]
Although the limiting behavior of $x_{\mathrm{min}}$ is left an
undeveloped problem in CW, the solution of the above optimization
problem could be a possible option heuristically. Numerical
optimization (grid search in matlab) yields $\alpha=2.31$ and $x_{\mathrm
{min}}=13.44$. With these values, the value of $p$ in equation (\ref
{eq:p}) is $0.0242$.

Table~\ref{default} reports the results for the bootstrap confidence intervals for
both the probability $p$ of a catastrophic event and $\alpha$, the
power-law parameter. In brief, out of the six different setups, the
confidence intervals failed for two of them. Under both (1) the mixed
power-law distribution [from \citet{clausetPower}] with $x_{\min}$
given and (2) the GPD with $x_{\min}$ estimated, the nominally $90\%$
bootstrap confidence intervals cover the true values of $p$ with
probabilities $0.64$ and $0.71$, respectively. One reason the bootstrap
fails under the mixed power-law distribution is possibly due to an
observation in CW, that a fixed choice of $x_{\mathrm{min}}$
underestimates the uncertainty in $\hat p$ due to the tail's unknown
structure. 
The reason the bootstrap fails under the GPD is that the algorithm
tends to underestimate $x_{\mathrm{min}}$ and $\alpha$, that is, it is
inclined for heavier tails. This is consistent with the other
discussants who suggest that $x_{\mathrm{min}}$ is potentially too small.

Although the bootstrap gives a straightforward path to computing
confidence intervals, these simulations suggest that their coverage
performance is sensitive to the data-generating model and whether or
not $x_{\min}$ is estimated or known.

%





\printaddresses

\end{document}